\documentclass[
twocolumn,
preprintnumbers,
nofootinbib,
showpacs,
prd,aps
]{revtex4}

\usepackage{graphicx}
\usepackage{amsmath, amssymb}

\begin{document}

\preprint{CALT 68-2860}

\title{
Dark Matter in Classically Scale-Invariant Two Singlets Standard Model 
}

\author{Koji Ishiwata}

\affiliation{
California Institute of Technology, Pasadena, CA 91125, USA
}

\date{December 12, 2011}

\begin{abstract}
  We consider a model where two new scalars are introduced in the
  standard model, assuming classical scale invariance. In this model
  the scale invariance is broken by quantum corrections and one of the
  new scalars acquires non-zero vacuum expectation value (VEV), which
  induces the electroweak symmetry breaking in the standard model, and
  the other scalar becomes dark matter.  It is shown that TeV scale
  dark matter is realized, independent of the value of the other
  scalar's VEV. The impact of the new scalars on the Higgs potential
  is also discussed. The Higgs potential is stabilized when the Higgs
  mass is over $\sim$120 GeV.
\end{abstract}

\maketitle

The hierarchy problem is one of the unsolved problems in standard
model of particle physics. In the standard model, Higgs boson is the
only scalar and it has a mass scale.  This mass is highly unstable if
radiative corrections are taken into account, which means that it must
be fine-tuned for the electroweak (EW) symmetry breakdown to occur at
$O(100~{\rm GeV})$.  Even if one tolerates the fine-tuning, the Higgs
sector has another problem. Higgs quartic coupling might become large
and have Landau pole or get negative at high scale below the Planck
scale \cite{Cabibbo:1979ay,Beg:1983tu,Lindner:1985uk,
  Altarelli:1994rb,Casas:1994us,Casas:1994qy,Hambye:1996wb}.  In order
to avoid a Landau pole up to the Planck scale and make the EW vacuum
cosmologically stable, the Higgs mass has to be about $115~{\rm
  GeV}$--$180~{\rm GeV}$ \cite{Hambye:1996wb,Ellis:2009tp}.
Otherwise, there should be new physics below the Planck scale. High energy
collider experiments are aimed at searching for the Higgs boson.  The
LEP experiments give a lower bound on the Higgs mass as $m_h\ge
114.4~{\rm GeV}$ ($m_h$ is standard-model Higgs mass)
\cite{Barate:2003sz}.  The region $156~{\rm GeV}\le m_h \le 177~{\rm
  GeV}$ is excluded by the Tevatron experiment \cite{arXiv:1108.3331}.
At the LHC, most of the region in $145~{\rm GeV}\le m_h\le 466~{\rm
  GeV}$ is excluded \cite{HiggsLHC}.

The existence of dark matter in the universe is another mystery to be
answered. Dark matter (DM) accounts for about five times of the baryon
energy density in the universe according to the recent observations
\cite{arXiv:1001.4538}. However, the standard model does not have a
good candidate for DM.

Probably most popular and elegant solution to the both problems is
supersymmetry. In a supersymmetric model, the fine-tuning for the
Higgs mass parameter is avoided when the mass of superparticles are
below about a TeV.  Moreover, since the Higgs quartic coupling is
written in terms of gauge coupling constants squared, it never has a
Landau pole or gets negative up to the Planck scale.  In addition, the
lightest superparticle (LSP) is stable under $R$-parity.  Such a
stable LSP, with a mass of the order a TeV, can be the dark matter,
which is the so-called weakly interacting massive particle (WIMP)
scenario.  However, the recent null result of the search of the
superparticles at the LHC \cite{Aad:2011ib,Chatrchyan:2011zy}
indicates that typical scale of superparticle mass should be above
around TeV. This result causes a tension in supersymmetric model in
terms of hierarchy problem.

Another approach for the solution of the hierarchy problem is to
consider scale invariance in theory. Even if a theory has scale
invariance classically, the scale invariance would be broken by
quantum corrections in Coleman-Weinberg mechanism
\cite{Coleman:1973jx}. Then a mass scale for scalar field is induced
(known as dimensional transmutation). This mass is protected from
quantum corrections due to the scale invariance at classical level
although the scale invariance is anomalous
\cite{FERMILAB-CONF-95-391-T,arXiv:0709.2750}. However, possibility of
Coleman-Weinberg mechanism in the standard-model Higgs sector is
already excluded due to large top Yukawa coupling.  In the framework
of scale-invariant theory, several simple extensions of the standard
model are proposed recently.  In Refs.~\cite{Iso:2009ss,Iso:2009nw}, a
new scalar charged with non-zero $B-L$ ($B$ and $L$ are baryon and
lepton numbers) breaks $U(1)_{B-L}$ gauge symmetry radiatively and
then the braking induces the EW symmetry breaking. Another extension
is adding new scalar singlets to the standard
model~\cite{arXiv:1006.0131,arXiv:1111.0273}. It is described in their
works, however, that dark matter can not be explained without
considering additional fields.\footnote
{It is mentioned in Refs.~\cite{Iso:2009ss,Iso:2009nw} an additional
  scalar is needed for dark matter candidate, while a scenario where
  dark matter in mirror standard model is considered in
  Ref.~\cite{arXiv:1006.0131}. In Ref.~\cite{arXiv:1111.0273}, heavy
  right-handed neutrino is discussed as a candidate for dark matter
  and further cosmological consequence of the scale-invariant theory
  is given.}
Scale-invariant two Higgs doublet model is studied in
Ref.~\cite{Hambye:2007vf}. In their work EW symmetry breaking is
induced due to more degrees of freedom of scalars, and one of the
neutral component is a candidate for dark matter. However, in their
scenario, the couplings are not perturbative below the Planck scale.
Also there is a study which proposes a dark matter candidate in the
scale-invariant extension with a strongly interacting hidden
sector~\cite{Hur:2011sv}.

In this Letter, we consider two new scalar singlets in classically
scale-invariant standard model, and study a possibility that one of
the scalars becomes dark matter. The mass of the scalar dark matter is
provided by the non-zero vacuum expectation value (VEV) of the other
scalar. The non-zero VEV is induced by quantum corrections to break
classical scale invariance, and then the EW symmetry is broken as
shown in Ref.~\cite{arXiv:1006.0131}. In the analysis we demand
perturbativity for all the couplings in the model up to the Planck
scale.  It is shown that the singlet can explain the present energy
density of dark matter with a mass of TeV for various values of the
VEV. It turns out that the scalar which has the non-zero VEV is
sequestered from the other sectors, then the properties of the singlet
DM becomes similar to the one in the standard model with a singlet
scalar \cite{DOE-ER-40048-13
  P5,hep-ph/0702143,hep-ph/0011335,Davoudiasl:2004be}. The impact of
the new scalars on Higgs potential is also discussed.  The Higgs lower
mass bound from the stability of the Higgs potential gets smaller and
it turns out to be $120$--$125~{\rm GeV}$ due to the existence of the
new scalars.

Let us start off with the framework of our model. We consider the
theory which has scale invariance at classical level, {\it i.e.}
there is no mass term in Lagrangian. In addition to the Higgs boson,
we introduce two real new standard-model singlet scalars, $\phi_1$ and
$\phi_2$, in scalar sector. Then, the renormalizable potential of the
scalar sector is written as
\begin{eqnarray}
V&=&\frac{1}{8}\lambda_1 \phi_1^4+\frac{1}{8}\lambda_2 \phi_2^4
+\frac{1}{4}\kappa_{12} \phi_1^2 \phi_2^2
+\frac{1}{2}\lambda_H (H^{\dagger}H)^2 
\nonumber \\
&+& \frac{1}{2}\kappa_{H1} H^{\dagger}H \phi_1^2
+ \frac{1}{2}\kappa_{H2} H^{\dagger}H \phi_2^2,
\end{eqnarray}
where $H$ is Higgs doublet.\footnote
{In Ref.~\cite{Abada:2011qb}, similar potential but with mass terms is
  discussed. In the work one of the singlets is a candidate for dark
  matter.}
Here we assume $Z_2$ symmetry, {\it i.e.}  the potential is invariant
under $\phi_2 \rightarrow -\phi_2$.  Then a term, such as
$\phi_1\phi_2 H^{\dagger} H$, is forbidden. As to couplings, we
consider all the couplings are positive, except for
$\kappa_{H1}$. However, a sizable negative value for $\kappa_{H1}$
might cause instability of the potential. Thus, in this Letter, we
assume very small $|\kappa_{H1}|$. In addition, we limit our study in
the region where all the couplings are under control in perturbation
up to the Planck scale.

First let us focus on the new scalars. Even $\kappa_{12}>0$, the
spontaneous breakdown of scale invariance would be induced by quantum
corrections.  Either $\phi_1$ or $\phi_2$ could have a VEV.  For the
EW symmetry breakdown in the standard model sector (by the negative
$\kappa_{H1}$), we consider the case where $\phi_1$ gets the non-zero
VEV.  The evaluation of the VEV is performed by the use of
renormalization group (RG) improved potential at one-loop level
\cite{Sher:1988mj}:
\begin{eqnarray}
V =\frac{1}{8}\lambda_1(t) \phi_1^4+\frac{1}{8}\lambda_2(t) \phi_2^4
+\frac{1}{4}\kappa_{12}(t) \phi_1^2 \phi_2^2+{\cdots}, 
\end{eqnarray}
where $t=\log(\mu/M)$.  Here $\mu$ is renormalization scale and $M$ is
arbitrary scale. The RG equations at one-loop level for $\lambda_1$,
$\lambda_2$ and $\kappa_{12}$ are given by
\begin{eqnarray}
16 \pi^2 \frac{d\lambda_1}{dt}
&=& 9 \lambda_1^2 + \kappa_{12}^2+4\kappa_{H1}^2, \label{eq:RGE1} \\
16 \pi^2 \frac{d\lambda_2}{dt}
&=& 9 \lambda_2^2 + \kappa_{12}^2+4\kappa_{H2}^2, \\
16 \pi^2 \frac{d\kappa_{12}}{dt}
&=& 4 \kappa_{12}^2 + 3 \kappa_{12} (\lambda_1+\lambda_2)
+4\kappa_{H1}\kappa_{H2}.
\label{eq:RGE3}
\end{eqnarray}
The RG equations for the other couplings are given in
Appendix. Hereafter we take $M=\langle \phi_1 \rangle$. ($\langle
\phi_1 \rangle$ denotes the VEV of $\phi_1$.) For the evaluation to
find stationary point where $\phi_1$ has the VEV (and $\langle \phi_2
\rangle$=0), taking $\mu=\phi_1$ is a good approximation.  Then
$\partial V/\partial\phi_1|_{\phi_1=\langle \phi_1 \rangle}=0$ gives a
condition,
\begin{eqnarray}
\left[\frac{d\lambda_1}{dt}+4\lambda_1\right]\bigg|_{t=0}=0,
\end{eqnarray}
and it leads to 
\begin{eqnarray}
\nonumber \\
\lambda_1(0)\simeq 
-\frac{1}{64 \pi^2}(\kappa_{12}^2(0)+4\kappa_{H1}^2(0)).
\label{eq:lambda1_ini}
\end{eqnarray} 
Consequently, the mass parameter of $\phi_1$ is calculated by the
second derivative of the potential with respect to $\phi_1$ at
$\phi_{1}=\langle \phi_1 \rangle$:
\begin{eqnarray}
m^2_{\phi_1} &=& \frac{\partial^2 V}{\partial^2 \phi_1} \bigg|_{t=0}
\nonumber \\
&\simeq& - 2 \lambda_1(0) M^2
\nonumber \\
&\simeq& \frac{1}{32\pi^2}(\kappa_{12}^2(0)+4\kappa_{H1}^2(0)) M^2.
\end{eqnarray}
Therefore, the negative value of $\lambda_1(0)$ gives the right sign
for the mass term of $\phi_1$.  One may wonder the negative
$\lambda_1(0)$ would induce a deeper minimum at some higher
scale. However, since $|\lambda_1(0)|$ is required to be small enough
(see Eq.~(\ref{eq:lambda1_ini})), the potential is stabilized.

The important aspect of the spontaneous symmetry breaking in $\phi_1$
sector is that the symmetry breaking also induces the EW symmetry
breakdown in the Higgs sector. By replacing $\phi_1$ as $M+\phi_1$ in
the potential, relevant terms for the EW symmetry breaking are given
by
\begin{eqnarray}
V = \frac{1}{2}\lambda_H (H^{\dagger}H)^2 
+  \frac{1}{2}\kappa_{H1}  M^2 H^{\dagger}H + \cdots. 
\end{eqnarray}
Then it is seen that the EW symmetry is broken (by the negative
$\kappa_{H1}$).  In the evaluation of the Higgs VEV, taking
renormalization scale to be the Higgs field is a good
approximation. Then $\lambda_{H}$ and $\kappa_{H1}$ relates at the
scale of $\mu=v$ ($v\simeq 246~{\rm GeV}$) as
\begin{eqnarray}
\mu_h^2(t_v) = -\kappa_{H1}(t_v) M^2.
\label{eq:kappa_{H1}}
\end{eqnarray}
Here $\mu^2_H(t)= \lambda_H(t) v^2$ and $t_v=\log(v/M)$. 

Due to the VEVs of $\phi_1$ and Higgs, both fields mix with each
other. Pluging $H=(v+h)/\sqrt{2}$ ($h$ is Higgs field) in the
potential, the mass terms for the mixed states are
\begin{eqnarray}
V^{\rm mass}_{h\phi_1}=\frac{1}{2}
\begin{array}{cc}
(\phi_1 & h)
\end{array}
\left( \begin{array}{cc}
m_{\phi_1}^2 & \Delta m^2 \\
\Delta m^2  & \mu_h^2 
\end{array}\right)
\left(\begin{array}{c}
\phi_1 \\ 
h
\end{array}\right),
\end{eqnarray}
where $\Delta m^2= \kappa_{H1}vM$. Then the mass eigenstates, denoted
as $s_a$ and $s_b$, are parametrized as
\begin{eqnarray}
\left(\begin{array}{c}
s_a \\ 
s_b
\end{array}\right)
=
\left( \begin{array}{cc}
\cos \theta_s & \sin \theta_s \\
 -\sin \theta_s  & \cos \theta_s
\end{array}\right)
\left(\begin{array}{c}
\phi_1 \\ 
h
\end{array}\right).
\end{eqnarray}
Here and hereafter we take $m_{s_a}<m_{s_b}$ ($m_{s_a}$ and $m_{s_a}$
are masses of $s_a$ and $s_b$, respectively) without loss of
generality.  When $m_{s_a}<114.4~{\rm GeV}$, the mass of $s_a$ would
be constrained by the LEP experiment.  The constraints are given by in
terms of $\xi^2=(g_{sZZ}/g_{HZZ}^{\rm SM})$ ($g_{sZZ}$ is new
scalar-$Z$-$Z$ coupling and $g_{HZZ}^{\rm SM}$ is $h$-$Z$-$Z$ coupling
in the standard model)~\cite{Barate:2003sz}.  In our case it is equal to
$\sin^2\theta_s$ and the scalar which has $\xi \gtrsim 0.01$ is
constrained when it is lighter than $114.4~{\rm GeV}$. Thus we
calculate $\xi^2$ to evaluate the constraints by the LEP.  On the
other hand, $\phi_2$ does not mix with the other scalars. Its mass is
induced by the $\langle \phi_1 \rangle$, and it is easily obtained as
\begin{eqnarray}
m_{\phi_2}^2 =\frac{1}{2}\kappa_{12}(0)M^2+\frac{1}{2}\kappa_{H2}(0)v^2.
\label{eq:mphi2}
\end{eqnarray}
Therefore, $m_{\phi_2}\gg m_{s_a},~m_{s_b}$ is expected in our
scenario.

\begin{widetext}

\begin{figure}[t]
  \begin{center}
    \includegraphics[scale=0.6]{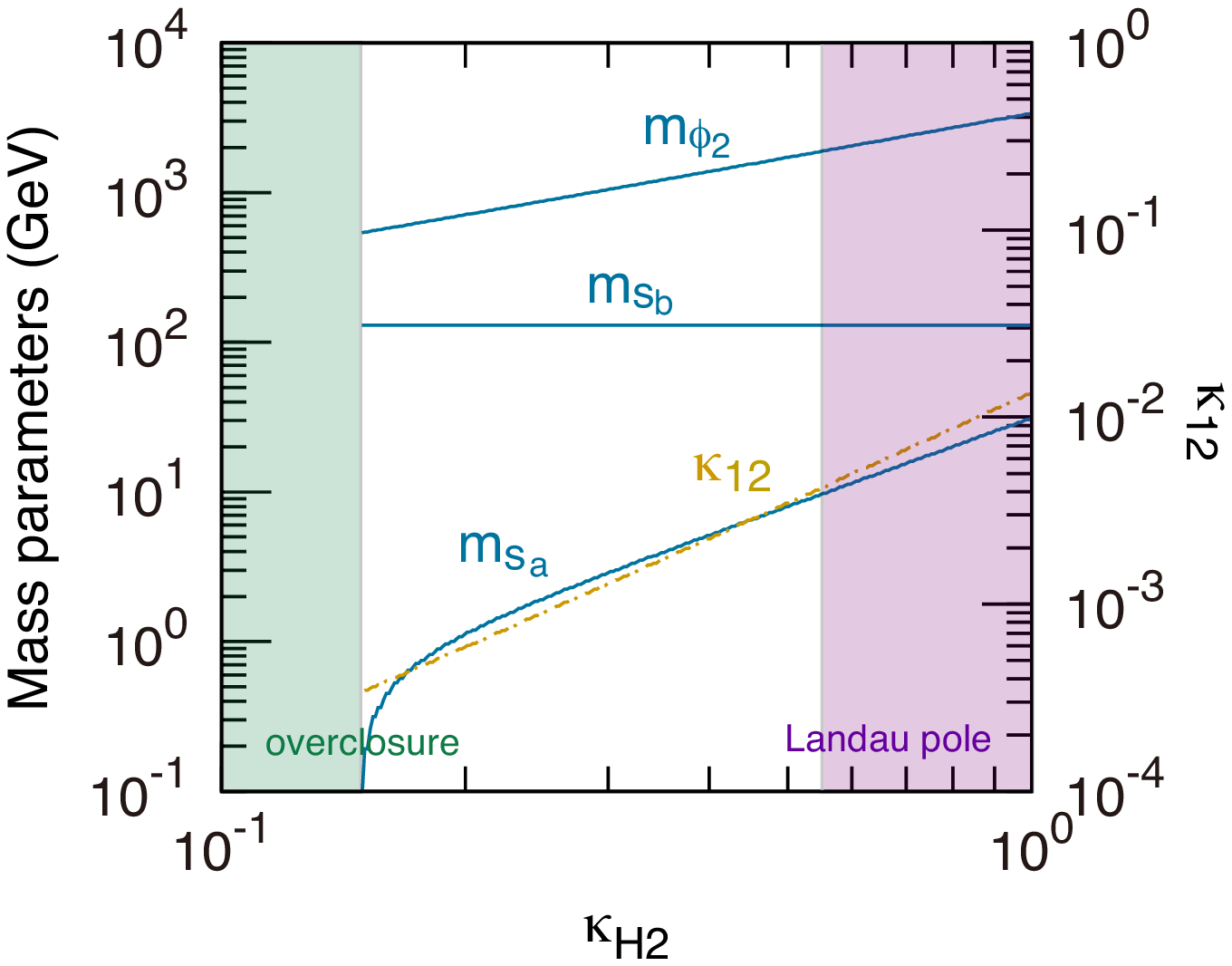}
    \includegraphics[scale=0.6]{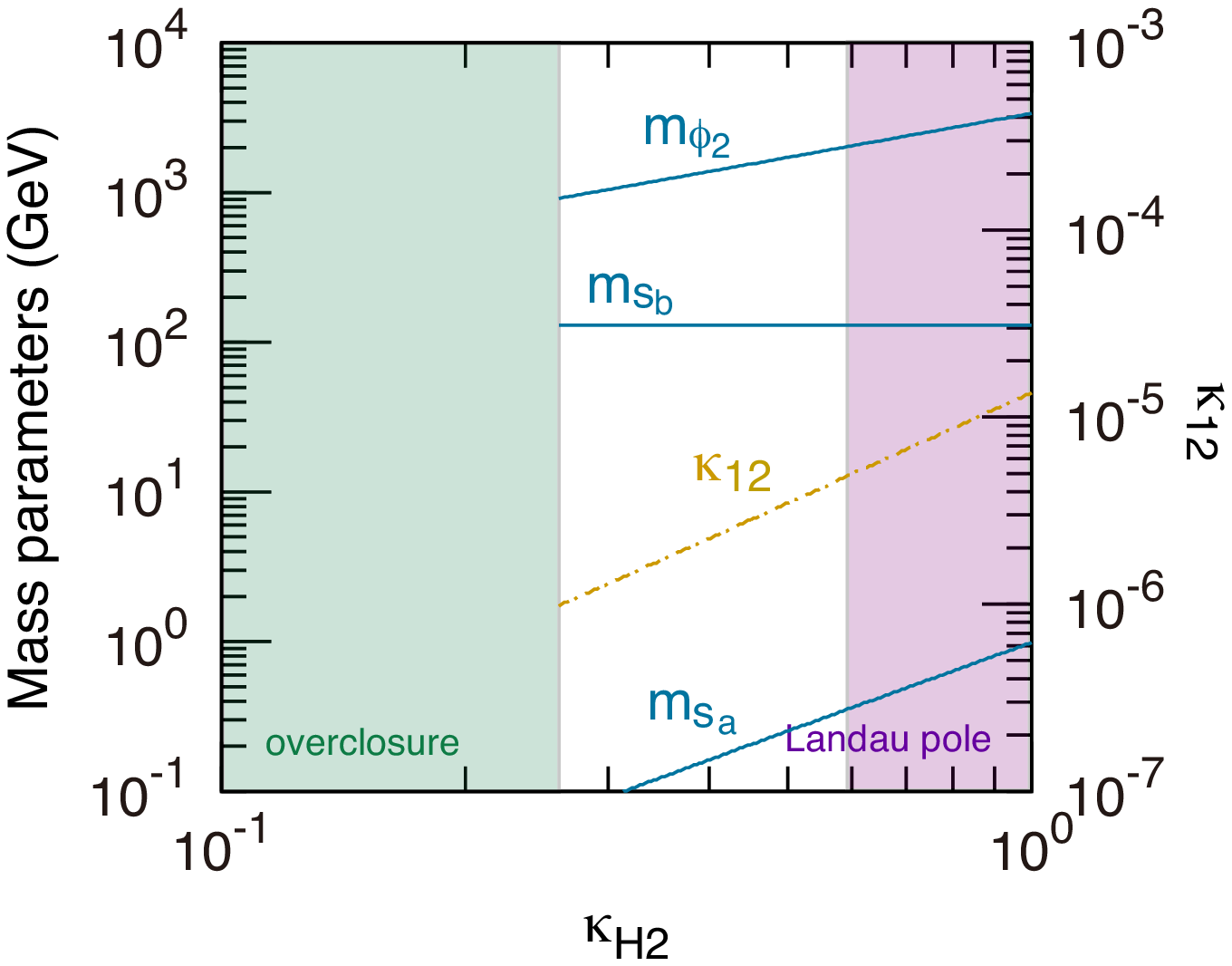}
  \end{center}
  \caption{Masses of $s_a$, $s_b$ and $\phi_2$ (left vertical axis)
    and $\kappa_{12}$ (right vertical axis) as the function of
    $\kappa_{H2}(\mu=m_{\phi_2})$. Here we take $\mu_h(t_v)=130~{\rm
      GeV}$ and $|\kappa_{H1}(t_v)|=10^{-5}$ (left panel), $10^{-8}$
    (right panel).  Light green shaded region shows that the lightest
    scalar becomes tachyon, while in dark purple shaded region
    $\kappa_{H2}$ has Landau pole.}
  \label{fig:spectramh130}
\end{figure}

\end{widetext}

Now we are ready to discuss phenomenological aspects of our
model. Since the $Z_2$ symmetry for $\phi_2$ is unbroken, $\phi_2$ is
stable and becomes a good candidate for dark matter.  In the early
universe, $\phi_2$ is produced thermally and the relic abundance is
determined by its annihilation cross section.  The annihilation cross
section depends on $\kappa_{12}$, $\kappa_{H2}$ and $\langle \phi_1
\rangle$. Relevant processes are $\phi_2 \phi_2 \rightarrow s_a s_a$,
$s_b s_b$, $s_a s_b$, $WW$, $ZZ$ and $f\bar{f}$. (Here $f$ stands for
standard-model fermion.)  When $\kappa_{H2}$ is negligible, main
process which contributes the annihilation is $\phi_2 \phi_2$ to the
scalar pairs.  In such a case, however, the annihilation cross section
is too small, which leads to overclosure of the universe. Therefore we
consider finite value of $\kappa_{H2}$, which allows the annihilation
channels to the standard-model gauge bosons and fermion pairs.  In our
model, when $\kappa_{H1}$, $\kappa_{H2}$ and $\kappa_{12}$ are fixed
for a given value of $\mu_h$, all the mass parameters are
determined. Then the annihilation cross section and the relic
abundance can be calculated. Therefore we impose
$\Omega_{\phi_2}=\Omega_{\rm DM}$ ($\Omega_{\phi_2}$ and $\Omega_{\rm
  DM}$ are the energy densities of $\phi_2$ and dark matter,
respectively) in our numerical calculation.

The mass spectra to realize $\phi_2$-DM scenario are given in
Fig.~\ref{fig:spectramh130}.  Here we take $\mu_h(t_v)=130~{\rm GeV}$,
and $|\kappa_{H1}(t_v)|=10^{-5}$ and $10^{-8}$ are taken in left and
right panels, respectively. Light shaded region is forbidden because
$s_a$ becomes tachyonic, while $\kappa_{H2}$ has a Landau pole in dark
shaded region.  In the evaluation, we have solved the RG equations
given in Eqs.~(\ref{eq:RGE1})-(\ref{eq:RGE3}) and
(\ref{eq:RGEH1})-(\ref{eq:RGEH}) for $\mu\ge m_{\phi_2}$. When $\mu
\le m_{\phi_2}$, we consider the effective theory in which $\phi_2$ is
integrated out. Thus, for $\lambda_H$ and $\kappa_{H1}$, we have
solved the RG equations given in Eqs.~(\ref{eq:RGE_lambdaHeff}) and
(\ref{eq:RGE_kH1eff}). (The RG evolution of $\lambda_H$ at low energy
is important to determine the Higgs pole mass. See later discussion.)
Here we note that horizontal axis in the Figure should be interpreted
as a value at the scale of $\mu=m_{\phi_2}$ because energy scale of
annihilation process is characterized by $\phi_2$ mass. Meanwhile, we
have checked running effect of $\kappa_{12}$ is very small. Thus we
have used values of $m_{\phi_1}$ and $m_{\phi_2}$ at $\mu=M$ for the
calculation of the pole masses. $\Delta m$ is evaluated by using
$\kappa_{H1}(t_v)$. In the allowed region, the mass of $\phi_2$-DM
turns out to be a few hundred GeV to a few TeV. This result indicates
that $\phi_2$-DM annihilates into the standard-model particles via
$\kappa_{H2}\sim O(0.1$--$1)$. Then, as in the standard WIMP scenario,
particle with a mass of TeV scale give the right amount relic to
explain the present DM abundance. For the other scalars, we found that
$s_a \simeq \phi_1$ and $s_b\simeq h$.  The behavior of $m_{s_a}$ can
be understood as follows.  From Eqs.~(\ref{eq:kappa_{H1}}) and
(\ref{eq:mphi2}), $\kappa_{12}\sim 10^2\kappa_{H1}$ for $\mu_h\sim
100~{\rm GeV}$ and $m_{\phi_2}\sim 1~{\rm TeV}$. Then the mass of
$s_a$ is estimated as $m_{s_a}\simeq m_{\phi_1}\sim 10^3
\sqrt{|\kappa_{H1}|}~{\rm GeV}$, which roughly agrees with the
numerical result given in the Figure. It is seen that the mass of the
lightest scalar is far below of the LEP bound. However, we have
checked that $\xi^2$ for $s_a$ is much smaller than $0.01$. Therefore,
the LEP experiment does not exclude the lightest scalar. Finally as to
a magnitude of $\kappa_{H1}$, its upper bound is expect as
$|\kappa_{H1}|\lesssim O(10^{-3})$ from perturbativity, using
$\kappa_{12}\sim 10^2 |\kappa_{H1}|\lesssim O(0.1)$. On the other
hand, lower bound is implicitly given by $|\kappa_{H1}|\gtrsim
\mu^2_h/M^2_{\rm pl}$ ($M_{\rm pl}=2.4\times 10^{18}~{\rm GeV}$) from
Eq.~(\ref{eq:kappa_{H1}}).

Since $|\kappa_{H1}|$ is very small but $\kappa_{H2}$ is sizable, the
properties of $\phi_2$-DM is similar to those of a singlet DM in new
minimal standard model (NMSM) which is studied in
Refs.~\cite{DOE-ER-40048-13
  P5,hep-ph/0702143,hep-ph/0011335,Davoudiasl:2004be}. Phenomenology
of the NMSM has been studied by many works up to the present. It is
pointed out possibility of direct detection of singlet DM. (See
Ref.~\cite{arXiv:1001.0486} for the updated analysis, in which DM with
a mass of less than $200~{\rm GeV}$ is considered.)  We calculated
spin independent cross section of $\phi_2$-DM with nucleon by
following \cite{arXiv:1007.2601} to include gluon contribution with
QCD correction. Then the cross section turns out to be about
$10^{-45}~{\rm cm}^2$ for $m_h=125~{\rm GeV}$.  The result has less
dependence on $\kappa_{H2}$ in the region where $|\kappa_{H1}|\lesssim
10^{-5}$. Thus it may be difficult to detect $\phi_2$-DM in the near
future direct detection experiments.

The new singlet scalars also have impact on the RG evolution of the
Higgs quartic coupling. In the standard model, $\lambda_{H}$ gets
negative below the Planck scale for $\mu_h(t_v)\lesssim 135~{\rm GeV}$
at the one-loop level RG equations.  In careful evaluation of the
Higgs pole mass at two-loop level, it is shown the Higgs pole mass
should be larger than about $130~{\rm GeV}$ in order for the Higgs
potential to be stable up to the Planck scale
\cite{Casas:1994us,Casas:1994qy}.\footnote
{When $m_h\lesssim 130~{\rm GeV}$, the Higgs potential has another
  deeper minimum below the Planck scale. However, the lifetime of the
  EW vacuum is longer than the age of universe when $m_h\gtrsim
  115~{\rm GeV}$ \cite{Ellis:2009tp}. }
In the recent work the Higgs lower mass bound in the NMSM is
calculated carefully at one-loop level. It is shown in
Ref.~\cite{arXiv:0910.3167} that the Higgs lower mass bound turns out
to be around 130 GeV in the NMSM.  In order to see the impact of the
new scalars to the Higgs potential in our model, we solved the RG
equations given in Eqs.~(\ref{eq:RGE1})-(\ref{eq:RGE3}) and
(\ref{eq:RGEH1})-(\ref{eq:RGEH}) for $\mu\ge m_{\phi_2}$ and
Eqs.~(\ref{eq:RGE_lambdaHeff}) and (\ref{eq:RGE_kH1eff}) for $\mu \le
m_{\phi_2}$.  In the present scenario sizable $\kappa_{H2}$ tends to
increase $\lambda_{H}$, then it turns out that $\lambda_{H}$ is
positive up to the Planck scale when $\mu_H(t_v)\gtrsim 120~{\rm
  GeV}$, taking possible largest value for $\kappa_{H2}$.  Following
the procedure given in Ref.~\cite{Casas:1994us}, we calculated the
Higgs pole mass at one-loop level,\footnote{In
  Ref.~\cite{Casas:1994us}, they solve RG equations at two-loop
  level. } and obtained $m_h\simeq 123~{\rm GeV}$ for $\mu_h=120~{\rm
  GeV}$. This means that the Higgs potential is stabilized up to the
Planck scale when the Higgs pole mass is larger than about $120~{\rm
  GeV}$. However, since this result is based on the one-loop RG
calculation, the lower Higgs mass bound might be changed by a few GeV
in more accurate calculation of the pole mass. Such an analysis is
beyond the scope of this Letter. It will be given elsewhere.

Finally we note on cosmological problem in our model. Since there is
also $Z_2$ symmetry under transformation, $\phi_1 \rightarrow
-\phi_1$, domain wall is formed as a topological defect when the
discrete symmetry breaks spontaneously. If the domain wall is stable
or long-lived, it would affect the primordial density fluctuation in
the early universe to give unobserved anisotropy in cosmic microwave
background (CMB). However, such a problem might be avoided if one
consider higher-dimensional operator which is suppressed by the Planck
mass which explicitly breaks the $Z_2$ symmetry, {\it e.g.},
$\phi_1^5/M_{\rm pl}$, assuming that scale invariance is broken at the
Planck scale. Then, two degenerate vacua splits to true and false
vacua. The energy density of the domain wall is roughly estimated as
$\rho_{\rm DW}\sim \sigma/R$ where $\sigma$ is the tension of the
domain wall and $R=M_{\rm pl}/T T_{\rm sb}$ is the scale of the domain
wall. Here $T$ is temperature and $T_{\rm sb}$ is the temperature at
the discrete symmetry breaking \cite{IASSNS-HEP-91-11}. If the energy
difference is larger than that of the domain wall, the false vacuum is
destroyed by the true vacuum before the domain wall dominates the
energy density of the universe. Consequently, the anisotropy in the
CMB is avoided. In our case $\sigma\sim M^3$ and $T_{\rm sb}~\sim
M$. Thus at the temperature $\rho_{\rm DW} \sim \rho_{R}(\sim T^4)$,
$\rho_{\rm DW}$ would be $\sim (M^4/M_{\rm pl})^{4/3}$. On the other
hand, the energy difference between true and false vacuum is
$M^5/M_{\rm pl}$. Therefore, the requirement for avoiding the domain
wall problem is sufficiently satisfied in our model. Another way to
avoid the domain wall problem is to consider reheating temperature
less than the VEV of $\phi_1$. Then the large entropy production at
the reheating dilutes the domain wall.

In conclusion we have studied classically scale-invariant standard
model with the new scalar singlets, as a solution for the hierarchy
problem.  We demand that all the coupling constants in the model are
perturbative up to the Planck scale. In our model one scalar has the
VEV and breaks the scale invariance, then the EW symmetry breaking
occurs. Although this scalar mixes with the standard-model Higgs
boson, the mixing is so small that the Higgs phenomenology at the
collider is unchanged.  The other scalar acquires its mass from the
VEV and becomes a good candidate for dark matter.  It is found that
dark matter with a mass of a TeV scale is realized for various value
of the VEV.  This scalar has sizable coupling to the Higgs and it may
be possible to stabilize the Higgs potential when the Higgs mass is
larger than $\sim$120 GeV.

\vspace{0.3cm}

\noindent
{\it Acknowledgements:}
The author is grateful to Mark B. Wise for meaningful discussions and
careful reading of the manuscript. The author also thanks Norimi
Yokozaki for useful comments and discussions. The work was supported
in part by the U.S. Department of Energy under contract
No. DE-FG02-92ER40701, and by the Gordon and Betty Moore Foundation.

\vspace{0.4cm}

\appendix
\section*{Appendix}

In the appendix, we give the RG equations for $\kappa_{H1}$,
$\kappa_{H2}$ and $\lambda_{H}$ and the annihilation cross section of
$\phi_2$.

The RG equations for $\kappa_{H1}$, $\kappa_{H2}$ and $\lambda_{H}$
are obtained as 
\begin{eqnarray}
16 \pi^2 \frac{d\kappa_{H1}}{dt}
&=& 4\kappa_{H1}^2 +\kappa_{H2}\kappa_{12}+
\kappa_{H1}(2 \gamma_h+3\lambda_1+6\lambda_H),
\label{eq:RGEH1}
\nonumber \\ \\
16 \pi^2 \frac{d\kappa_{H2}}{dt}
&=& 4\kappa_{H2}^2 +\kappa_{H1}\kappa_{12}+
\kappa_{H2}(2 \gamma_h+3\lambda_2+6\lambda_H),
\nonumber \\ \\
16 \pi^2 \frac{d\lambda_{H}}{dt}
&=&12\lambda_H^2+4 \lambda_H \gamma_h -12y_t^4
\nonumber \\ &&
+\frac{3}{4}(g_1^4+2 g_1^2g_2^2+3g_2^4) 
+\kappa_{H1}^2+\kappa_{H2}^2.
\label{eq:RGEH}
\end{eqnarray}
Here $\gamma_h=-(9/4) g_2^2-(3/4)g_1^2+3y_t^2$ ($g_1$, $g_2$ and $y_t$
are $U(1)_Y$, $SU(2)_L$ gauge couplings and top Yukawa coupling,
respectively).  In the calculation, we also solve the RG equations for
gauge and top Yukawa couplings. In the evaluation of the running of
top Yukawa, we use initial condition, $y_t(\mu=m_t)=\sqrt{2}m_t
(1+4\alpha_s(m_t)/3\pi)^{-1}/v$ where $m_t=171~{\rm GeV}$ and
$\alpha_s(m_t)$ is strong coupling at the scale of $\mu=m_t$
\cite{Casas:1994qy}. For $\mu\le m_{\phi_2}$, $\phi_2$ is integrated
out. Since we need to know the running of $\lambda_H$ in this energy
region for the determination of the Higgs pole mass, we use RG equations:
\begin{eqnarray}
16 \pi^2 \frac{d\lambda_{H}}{dt}
&=&12\lambda_H^2+4 \lambda_H \gamma_h -12y_t^4
\nonumber \\ &&
+\frac{3}{4}(g_1^4+2 g_1^2g_2^2+3g_2^4) 
+\kappa_{H1}^2,
\label{eq:RGE_lambdaHeff}
\\
16 \pi^2 \frac{d\kappa_{H1}}{dt}
&=& 4\kappa_{H1}^2 +
\kappa_{H1}(2 \gamma_h+3\lambda_1+6\lambda_H).
\label{eq:RGE_kH1eff}
\end{eqnarray} 
Although we have solved the RG equation for $\kappa_{H1}$, it turns
out that  the influence of $\kappa_{H1}$ on $\lambda_H$ is very
small. Thus the running of $\lambda_H$ is almost the same as in the
standard model for $\mu \le m_{\phi_2}$.

The annihilation cross sections of $\phi_2$ for each channel, $\phi_2
\phi_2 \rightarrow s_a s_a$, $s_b s_b$, $s_a s_b$, $WW$, $ZZ$ and
$\bar{f}f$, are given as follows:
\begin{eqnarray}
  \sigma v({\phi_2\phi_2\rightarrow s_a s_a})
&=&\frac{\beta_i(m_{s_a},m_{s_a})}{64 \pi m_{\phi_2}^2}{\cal M}^2_{aa}, \\
  \sigma v({\phi_2\phi_2\rightarrow s_b s_b})
&=&\frac{\beta_i(m_{s_b},m_{s_b})}{64 \pi m_{\phi_2}^2}{\cal M}^2_{bb}, \\
  \sigma v({\phi_2\phi_2\rightarrow s_a s_b})
&=&\frac{\beta_i(m_{s_a},m_{s_b})}{32 \pi m_{\phi_2}^2}{\cal M}^2_{ab}, \\
 \sigma v({\phi_2\phi_2\rightarrow WW})
&=&\frac{g_2^2\beta_i(m_W,m_W)}{16 \pi m_{\phi_2}^2} {\cal M}^2
\nonumber \\   && \! \! \! \! \!  
\times \frac{m_W^2}{m_{\phi_2}^2}
\Bigl[1+\frac{1}{2}\Bigl(1- \frac{2m_{\phi_2}^2}{m_W^2}\Bigr)^2\Bigr]
 , \nonumber \\ \\
  \sigma v({\phi_2\phi_2\rightarrow ZZ})
&=&\frac{g_Z^2\beta_i(m_Z,m_Z)}{32 \pi m_{\phi_2}^2} {\cal M}^2 
\nonumber \\ && 　\! \! \! \! \! 
\times \frac{m_Z^2}{m_{\phi_2}^2}
\Bigl[1+\frac{1}{2}\Bigl(1- \frac{2m_{\phi_2}^2}{m_Z^2}\Bigr)^2\Bigr]
, \nonumber \\ \\
  \sigma v({\phi_2\phi_2\rightarrow f\bar{f}})
&=&\frac{y_f^2\beta_i(m_f,m_f)}{8 \pi m_{\phi_2}^2}
\Bigl(1-\frac{m^2_f}{m_{\phi_2}^2}\Bigr) {\cal M}^2 .
\nonumber \\
\end{eqnarray}
Here $g_Z=\sqrt{g_1^2+g_2^2}$, and $m_W$ and $m_Z$ are $W$ and $Z$
boson masses, respectively. $m_f$ is fermion mass and its Yukawa
coupling is given by $y_f=\sqrt{2}m_f/v$. $\beta_i$ is defined as
$\beta^2_i(m_1,m_2)=(s^2-2s(m_1^2+m_2^2)+(m_1^2-m_2^2)^2)/s^2$ with
$s=4 m_{\phi_2}^2$.  The matrix elements of the scattering, ${\cal
  M}_{aa}$, ${\cal M}_{bb}$ and ${\cal M}$, are obtained as
\begin{widetext}
\begin{eqnarray}
{\cal M}_{aa} &=&
-(\kappa_{12} \cos^2 \theta_s + \kappa_{H2} \sin^2 \theta_s )
+\frac{2(\kappa_{12} \cos \theta_s M+ \kappa_{H2} \sin \theta_s v )^2}
{2m_{\phi_2}^2-m_{s_a}^2}, \\
{\cal M}_{bb} &=&
-(\kappa_{12} \sin^2 \theta_s + \kappa_{H2} \cos^2 \theta_s )
+\frac{2(-\kappa_{12} \sin \theta_s M+ \kappa_{H2} \cos \theta_s v )^2}
{2m_{\phi_2}^2-m_{s_b}^2}
-\kappa_{H2} \lambda_H v^2
\Bigr[\frac{\cos^2 \theta_s\sin^2 \theta_s}{4 m^2_{\phi_2}-m^2_{s_a}}
+\frac{3\cos^4 \theta_s}{4 m^2_{\phi_2}-m^2_{s_b}} \Bigl], \\
{\cal M}_{ab} &=&
(\kappa_{12}- \kappa_{H2})\cos^2 \theta_s  \sin^2 \theta_s 
-
(\kappa_{12} \cos \theta_s M+ \kappa_{H2} \sin \theta_s v )
(-\kappa_{12} \sin \theta_s M+ \kappa_{H2} \cos \theta_s v )
\times
\Bigr[ \frac{1}{t-m_{\phi_2}^2}+ \frac{1}{u-m_{\phi_2}^2}\Bigl],
\nonumber \\
{\cal M} &=&
m_{\phi_2}\Bigl[\frac{(\kappa_{12} \cos \theta_s M+ \kappa_{H2} \sin \theta_s v ) 
\sin \theta_s }{4 m_{\phi_2}^2-m_{s_a}^2}
+
\frac{(-\kappa_{12} \sin \theta_s M+ \kappa_{H2} \cos \theta_s v )
\cos \theta_s}{4 m_{\phi_2}^2-m_{s_b}^2}
\Bigr].
\end{eqnarray}
Here $t=m_{\phi_2}^2+m_{s_a}^2-2 m_{\phi_2}(m_{s_a}^2
+m_{\phi_2}^2\beta^2_i(m_{s_a},m_{s_b}))^{1/2}$ and
$u=m_{\phi_2}^2+m_{s_b}^2-2 m_{\phi_2}(m_{s_b}^2+
m_{\phi_2}^2\beta^2_i(m_{s_a},m_{s_b}))^{1/2}$.

\end{widetext}



\begin{thebibliography}{99}

\bibitem{Cabibbo:1979ay}
  N.~Cabibbo, L.~Maiani, G.~Parisi and R.~Petronzio,
  Nucl.\ Phys.\  B {\bf 158}, 295 (1979).

\bibitem{Beg:1983tu}
  M.~A.~B.~Beg, C.~Panagiotakopoulos and A.~Sirlin,
  Phys.\ Rev.\ Lett.\  {\bf 52}, 883 (1984).

\bibitem{Lindner:1985uk}
  M.~Lindner,
  Z.\ Phys.\  {\bf C31}, 295 (1986).
 
\bibitem{Altarelli:1994rb}
  G.~Altarelli, G.~Isidori,
  Phys.\ Lett.\  {\bf B337}, 141-144 (1994).

\bibitem{Casas:1994us}
  J.~A.~Casas, J.~R.~Espinosa, M.~Quiros, A.~Riotto,
  Nucl.\ Phys.\  {\bf B436}, 3-29 (1995).
  [arXiv:hep-ph/9407389 [hep-ph]].  

\bibitem{Casas:1994qy}
  J.~A.~Casas, J.~R.~Espinosa, M.~Quiros,
  Phys.\ Lett.\  {\bf B342}, 171-179 (1995).
  [hep-ph/9409458].

\bibitem{Hambye:1996wb}
  T.~Hambye, K.~Riesselmann,
  Phys.\ Rev.\  {\bf D55}, 7255-7262 (1997).
  [hep-ph/9610272].

\bibitem{Ellis:2009tp}
  J.~Ellis, J.~R.~Espinosa, G.~F.~Giudice, A.~Hoecker, A.~Riotto,
  Phys.\ Lett.\  {\bf B679}, 369-375 (2009).
  [arXiv:0906.0954 [hep-ph]].

\bibitem{Barate:2003sz}
  R.~Barate {\it et al.} [ LEP Working Group for Higgs boson searches and ALEPH and DELPHI and L3 and OPAL Collaborations ],
  Phys.\ Lett.\  {\bf B565}, 61-75 (2003).
  [hep-ex/0306033].

\bibitem{arXiv:1108.3331} 
  D.~Benjamin [for the CDF and D0 and The TEVNPH Working Group (The Tevatron New Phenomena and Higgs Working Group) Collaborations],
  arXiv:1108.3331 [hep-ex].

\bibitem{HiggsLHC}
  Based on the talk ``Higgs Searches at the LHC''
  given by Monica L. Vazquez Acosta at SUSY11, 2011

\bibitem{arXiv:1001.4538} 
  E.~Komatsu {\it et al.} [WMAP Collaboration],
  Astrophys.\ J.\ Suppl.\ \ {\bf 192}, 18  (2011)
  [arXiv:1001.4538 [astro-ph.CO]].

\bibitem{Aad:2011ib}
  G.~Aad {\it et al.}  [ATLAS Collaboration],
  arXiv:1109.6572 [hep-ex].

\bibitem{Chatrchyan:2011zy}
 S.~Chatrchyan {\it et al.} [CMS Collaboration],
 [arXiv:1109.2352 [hep-ex]].

\bibitem{Coleman:1973jx}
  S.~R.~Coleman, E.~J.~Weinberg,
  Phys.\ Rev.\  {\bf D7}, 1888-1910 (1973).

\bibitem{FERMILAB-CONF-95-391-T} 
  W.~A.~Bardeen,
  FERMILAB-CONF-95-391-T.

\bibitem{arXiv:0709.2750} 
  R.~Foot, A.~Kobakhidze, K.~L.~McDonald and R.~R.~Volkas,
  Phys.\ Rev.\ D\ {\bf 77}, 035006  (2008)
  [arXiv:0709.2750 [hep-ph]].


\bibitem{Iso:2009ss}
  S.~Iso, N.~Okada, Y.~Orikasa,
  Phys.\ Lett.\  {\bf B676}, 81-87 (2009).
  [arXiv:0902.4050 [hep-ph]].

\bibitem{Iso:2009nw}
  S.~Iso, N.~Okada, Y.~Orikasa,
  Phys.\ Rev.\  {\bf D80}, 115007 (2009).
  [arXiv:0909.0128 [hep-ph]].

\bibitem{arXiv:1006.0131} 
  R.~Foot, A.~Kobakhidze and R.~R.~Volkas,
  Phys.\ Rev.\ D\ {\bf 82}, 035005  (2010)
  [arXiv:1006.0131 [hep-ph]].

\bibitem{arXiv:1111.0273} 
  L.~Boyle, S.~Farnsworth, J.~Fitzgerald and M.~Schade,
  arXiv:1111.0273 [hep-ph].

\bibitem{Hambye:2007vf} 
  T.~Hambye and M.~H.~G.~Tytgat,
  Phys.\ Lett.\ B {\bf 659}, 651 (2008)
  [arXiv:0707.0633 [hep-ph]].

\bibitem{Hur:2011sv} 
  T.~Hur and P.~Ko,
  Phys.\ Rev.\ Lett.\  {\bf 106}, 141802 (2011)
  [arXiv:1103.2571 [hep-ph]].

\bibitem{DOE-ER-40048-13 P5} 
  V.~Silveira and A.~Zee,
  Phys.\ Lett.\ B\ {\bf 161}, 136  (1985).

\bibitem{hep-ph/0702143} 
  J.~McDonald,
  Phys.\ Rev.\ D\ {\bf 50}, 3637  (1994)
  [hep-ph/0702143 [HEP-PH]].

\bibitem{hep-ph/0011335} 
  C.~P.~Burgess, M.~Pospelov and T.~ter Veldhuis,
  Nucl.\ Phys.\ B\ {\bf 619}, 709  (2001)
  [hep-ph/0011335].

\bibitem{Davoudiasl:2004be}
  H.~Davoudiasl, R.~Kitano, T.~Li, H.~Murayama,
  Phys.\ Lett.\  {\bf B609}, 117-123 (2005).
  [hep-ph/0405097].

\bibitem{Abada:2011qb} 
  A.~Abada, D.~Ghaffor and S.~Nasri,
  Phys.\ Rev.\ D {\bf 83}, 095021 (2011)
  [arXiv:1101.0365 [hep-ph]].

\bibitem{Sher:1988mj}
  M.~Sher,
  Phys.\ Rept.\  {\bf 179}, 273-418 (1989).

\bibitem{KolbTurner}
 E.~W.~Kolb and M.~S.~Turner, `` The Early Universe'', Westview Press, 1990.

\bibitem{arXiv:0910.3167} 
  M.~Gonderinger, Y.~Li, H.~Patel and M.~J.~Ramsey-Musolf,
  JHEP\ {\bf 1001}, 053  (2010)
  [arXiv:0910.3167 [hep-ph]].

\bibitem{arXiv:1001.0486} 
  M.~Asano and R.~Kitano,
  Phys.\ Rev.\ D\ {\bf 81}, 054506  (2010)
  [arXiv:1001.0486 [hep-ph]].

\bibitem{arXiv:1007.2601} 
  J.~Hisano, K.~Ishiwata and N.~Nagata,
  Phys.\ Rev.\ D\ {\bf 82}, 115007  (2010)
  [arXiv:1007.2601 [hep-ph]].

\bibitem{IASSNS-HEP-91-11} 
  J.~Preskill, S.~P.~Trivedi, F.~Wilczek and M.~B.~Wise,
  Nucl.\ Phys.\ B\ {\bf 363}, 207  (1991).

\end{thebibliography}
\end{document}